\documentclass[twocolumn]{aastex631}
\pdfoutput=1
\hypersetup{linkcolor=red,citecolor=blue,filecolor=cyan,urlcolor=magenta}

\graphicspath{{./}{figures/}}

\submitjournal{ApJL}

\shorttitle{An unexpectedly bifurcated red-giant branch in a young massive star cluster}
\shortauthors{S. Kapse et al.}

\begin{document}

\title{Decoding the bifurcated red-giant branch as a tracer of multiple stellar populations in the young Large Magellanic Cloud cluster NGC 2173}

\correspondingauthor{Shalmalee Kapse}
\email{shalmalee.kapse@students.mq.edu.au}

\author{Shalmalee Kapse}
\affiliation{School of Mathematical and Physical Sciences, Macquarie University, Balaclava Road, Sydney, NSW 2109, Australia}
\affiliation{Research Centre for Astronomy, Astrophysics and Astrophotonics, Macquarie University, Balaclava Road, Sydney, NSW 2109, Australia}

\author{Richard de Grijs}
\affiliation{School of Mathematical and Physical Sciences, Macquarie University, Balaclava Road, Sydney, NSW 2109, Australia}
\affiliation{Research Centre for Astronomy, Astrophysics and Astrophotonics, Macquarie University, Balaclava Road, Sydney, NSW 2109, Australia}

\author{Devika Kamath}
\affiliation{School of Mathematical and Physical Sciences, Macquarie University, Balaclava Road, Sydney, NSW 2109, Australia}
\affiliation{Research Centre for Astronomy, Astrophysics and Astrophotonics, Macquarie University, Balaclava Road, Sydney, NSW 2109, Australia}

\author{Daniel B. Zucker}
\affiliation{School of Mathematical and Physical Sciences, Macquarie University, Balaclava Road, Sydney, NSW 2109, Australia}
\affiliation{Research Centre for Astronomy, Astrophysics and Astrophotonics, Macquarie University, Balaclava Road, Sydney, NSW 2109, Australia}




\begin{abstract}
Multiple stellar populations (MPs) representing star-to-star light-element abundance variations are common in nearly all ancient Galactic globular clusters. Here we provide the strongest evidence yet that the populous, $\sim$1.7 Gyr-old Large Magellanic Cloud cluster NGC 2173 also exhibits light-element abundance variations. Thus, our results suggest that NGC 2173 is the youngest cluster for which MPs have been confirmed to date. Our conclusion is based on the distinct bifurcation at the tip of its red-giant branch in high-quality color--magnitude diagrams generated from {\sl Hubble Space Telescope} imaging observations. Our results are further supported by a detailed analysis of `pseudo-$UBI$' maps, which reveal clear evidence of a bimodality in the cluster's red-giant-branch color distribution. Young clusters in the Magellanic Clouds can provide critical insights into galaxy evolution histories. Our discovery of MPs in NGC 2173 suggests that ancient Galactic globular clusters and young massive clusters might share a common formation process.
\end{abstract}

\keywords{Star clusters (1567) --- Hertzsprung Russell diagram (725) --- Large Magellanic Cloud (903) --- Red giant stars (1372) --- Stellar populations (1622)}


\section{Introduction} \label{sec:intro}

The consensus that all well-populated star clusters are `simple' stellar populations, fully described by a single isochrone is long gone. Modern observations have revealed that nearly all massive Galactic globular clusters (GCs) older than $\sim$6 Gyr are composed of multiple stellar populations (MPs). They often display multiple red-giant branches \citep[RGBs;][]{2009A&A...497..755M}, subgiant branches \citep[SGBs;][]{Marino2016ChemicalComponent}, and sometimes even multiple main sequences \citep[MSs;][]{Milone2016a}. This multiplicity is firmly associated with light-element abundance variations, a notion supported by spectroscopic observations. Spectroscopy of individual GC stars has yielded direct measurements of significant star-to-star abundance variations in chemical elements such as C, N, O, Na, Mg, and Fe \citep{Carretta2009IntrinsicClusters, Marino2008}. 

Many scenarios have been suggested to explain the observed multiplicity of features in color--magnitude space \citep[e.g.,]{Decressin2007OriginClusters,Bastian2013,DErcole2008FormationClusters}, but a self-consistent model that can explain all observational results remains elusive. In fact, recent studies have revealed that MPs are not only present in ancient Galactic GCs \citep{Milone2017} but also in populous Magellanic Cloud star clusters \citep{Mucciarelli2010ChemicalCloud, 2016ApJ...829...77D} with ages as young as $\sim$2 Gyr \citep{Niederhofer2017b, Martocchia2018AgeClusters}, as evidenced by multiple sequences in color--magnitude space \citep[e.g.,][]{2009A&A...497..755M}. However, the observed multiplicity of features in the color--magnitude diagrams (CMDs) of younger clusters is not usually caused by chemical abundance variations---as for old GCs---but may instead be related to stellar binarity and/or a range in stellar rotation rates.

Although both age and mass appear to correlate with the occurrence of MPs, only a few massive \citep[$\ga$ a few $\times 10^4 M_\odot$;][]{2009A&A...497..755M} clusters near the critical lower age limit of $\sim$2 Gyr have been studied in detail \citep[e.g.,][]{Kapse2021Searching2213}. It is as yet unclear which fundamental cluster parameter determines the onset of the occurrence of MPs. Securing a firm answer to this outstanding issue will offer novel insights into many key questions related to star cluster chemical enrichment histories and the star cluster--host galaxy connection. 

\begin{figure*}[ht]
    \begin{center}  
    \includegraphics[width=1\textwidth]{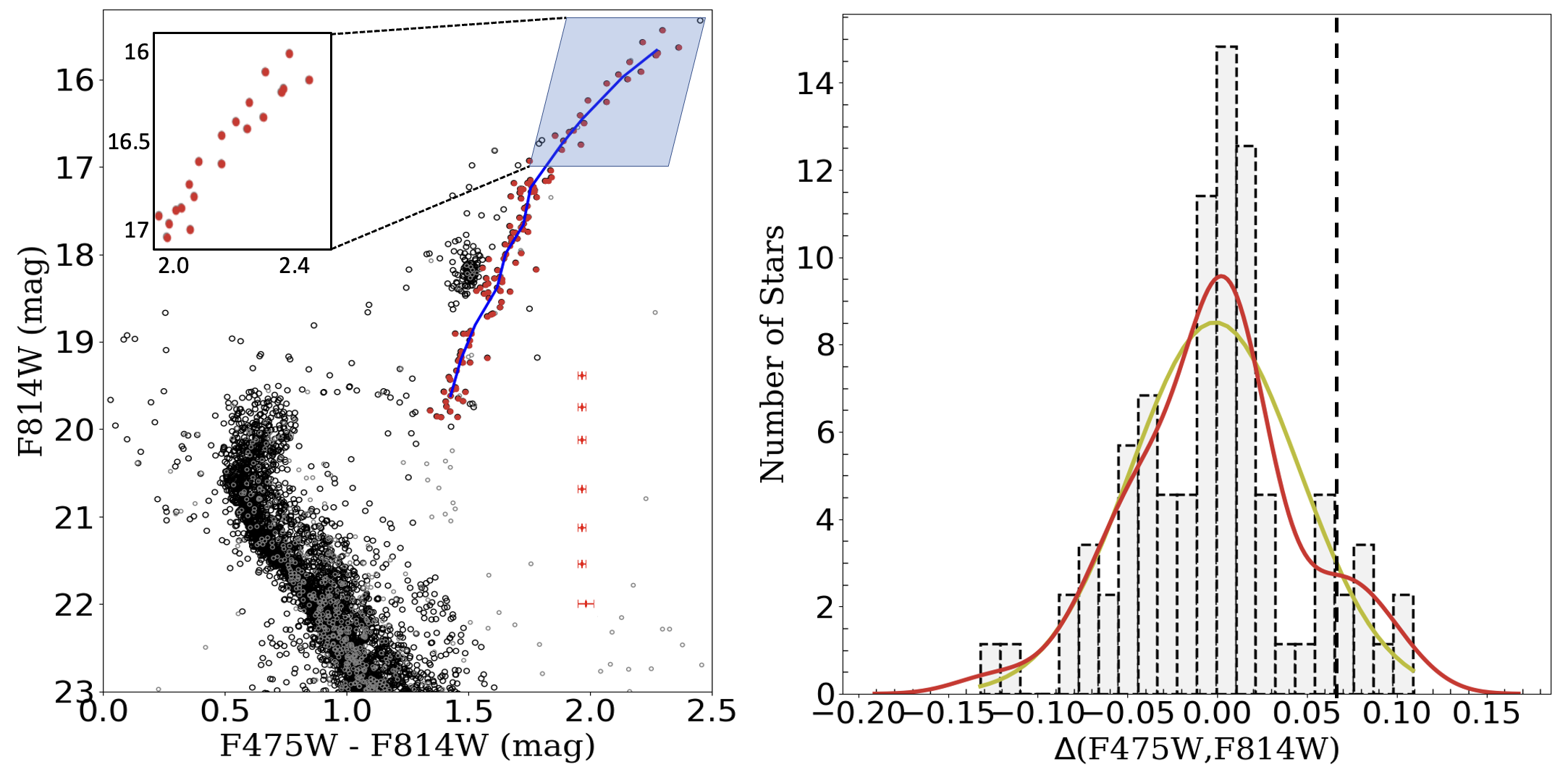}
    \end{center}
    \caption{(a) $(m_{\rm F475W} - m_{\rm F814W})$ versus $m_{\rm F814W}$ CMD of NGC 2173. Red dots: RGB stars; blue solid line: Fiducial line used for our RGB analysis; small gray dots: Field stars. Inset: Bifurcated TRGB region. (b) Verticalized $\Delta_{\rm F475W,F814W}$ RGB color distribution. Yellow, red solid lines: Gaussian probability distribution, kernel density estimation (KDE).}
    \label{fig1}
\end{figure*}

A number of systematic searches for MPs in Galactic and Magellanic Cloud clusters have recently been undertaken using {\sl Hubble Space Telescope} ({\sl HST}) imaging observations \citep{Piotto2015ThePopulations, Martocchia2017a, Martocchia2018a}. In this Letter, we present our analysis of ultraviolet (UV) and near-infrared {\sl HST} observations of NGC 2173, a young, $1.7 \pm 0.2$ Gyr-old Large Magellanic Cloud (LMC) cluster \citep{Glatt2008} with a mass of $\sim$$10^{5} M_\odot$ \citep[][]{2019ApJ...871..171L}. \citet{Li2018An} claimed to have detected a metallicity spread along its blue straggler sequence. However, using the same data, \citet{Dalessandro2019TheContamination} rejected a MP scenario. Here we show that the NGC 2173 RGB exhibits a clear, non-zero width, strongly suggesting a potential metallicity or light-element abundance spread.

\citet{Mucciarelli2008THE} analyzed spectra of four stars at the cluster's tip of its RGB (TRGB), but they did not find any evidence of chemical enrichment among their small sample. Based on photometric {\sl HST} observations, here we report the clear presence of MPs, using a significantly larger sample of 107 RGB stars. We also analyze a pseudo-color--color diagram resembling a `chromosome map' \citep{Milone2017}, henceforth referred to as a `pseudo-$UBI$ map' (see Section \ref{results.sec}). Chromosome maps (and our pseudo-$UBI$ map) are excellent tools for characterizing MPs in star clusters. At an age of 1.7 Gyr, NGC 2173 is now the youngest cluster where the presence of MPs has been confirmed, as we will show below. Despite the prevailing uncertainties, it is clearly younger than the previous youngest-age record holder, NGC 1978 \citep[$2.0 \pm 0.2$ Gyr;][]{Martocchia2018AgeClusters}. 

\section{OBSERVATIONS AND DATA REDUCTION}

Our analysis is based on data obtained with the {\sl HST}/Wide Field Camera 3 (WFC3) Ultraviolet--Visible channel (UVIS), available from the {\sl HST} Legacy Archive. We downloaded three sets of images in the F336W, F475W, and F814W passbands (program ID: GO-12257; PI: L. Girardi), with total exposure times of 2200 s, 1520 s, and 1980 s, respectively. These filters are approximately equivalent to the Johnson--Cousins $U, B$, and $I$ bands, respectively.

We applied point-spread function (PSF) photometry to the \texttt{.drz} and \texttt{.flt} data frames using the standard recipes implemented in the DOLPHOT2.0 photometry package \citep{Dolphin2011DOLPHOT/WFC32.0},\footnote{http://americano.dolphinsim.com/dolphot/dolphotWFC3.pdf} adopting the VegaMag photometric system. DOLPHOT was designed for analysis of {\sl HST} photometry. We used its UVIS/WFC3 modules (which include WFC3 PSFs and pixel area maps). They apply charge-transfer inefficiency corrections and include photometric calibration routines, such as zero-point and aperture corrections.

We obtained high-quality final stellar catalogs in all passbands. To select the `good' stars flagged by DOLPHOT, we only selected stars with `sharpness' values in the range $[-0.3, 0.3]$ and crowding $< 0.5$. The sharpness parameter compares a star's observed profile with the prevailing PSF; a perfect star has a sharpness of zero, a small negative sharpness value indicates a cosmic ray, and a large positive value implies that the detected object is extended or, most likely, a background feature such as a galaxy. DOLPHOT fits all stellar photometric profiles simultaneously. The crowding parameter, which quantifies the stellar brightness, is zero for isolated stars, while poorly measured stars generally return large values. 

This approach facilitates the removal of bad pixels, extended objects, cosmic rays, and objects affected by significant crowding. We thus selected, respectively, 150,723, 165,352, and 185,345 stars in our three UVIS/WFC3 frames.  

\subsection{Differential reddening correction}

First, we determined the cluster's center coordinates by superimposing number-density contour profiles onto the stellar spatial distribution, yielding $\alpha_{\rm J2000} = 05^{\rm h} 54^{\rm m} 17.63^{\rm s}, \delta_{\rm J2000} = -72^{\circ} 58' 38.46''$. These center coordinates closely match those of \citet{Li2018An}. To ensure selection of an RGB sample free from  field-star contamination, we first defined a region with a radius of $78''$, i.e., three times larger than the cluster's core radius \citep[]{Keller2012Extended2209}. 

We statistically subtracted the field stellar sample from the full photometric catalog encompassing the cluster region (see Figure \ref{fig1}, left: small gray dots). The decontaminated $(m_{\rm F475W} - m_{\rm F814W})$ versus $m_{\rm F475W}$ CMD is shown in Figure \ref{fig1} (left). The appearance of this and other NGC 2173 CMDs suggests that the cluster is not severely affected by differential reddening. Nevertheless, we corrected for differential reddening using the common approach illustrated by \citet{Milone2012}.Briefly, we rotated the CMD such that the vertical axis was oriented parallel to the direction of any differential reddening expected. We calculated the rotation angle using the ratio of the extinction coefficients in the filters used to generate the CMD. We next fitted a central, fiducial curve to our RGB sample and calculated the separation from this curve to each sample star. Finally, we converted this value into a star's local differential reddening, $\Delta E(B-V)$. The resulting reddening values were negligible; our corrections did not produce any significant changes in the cluster's CMDs.

\subsection{Selection of red-giant-branch stars}

Our aim is to analyze the cluster's decontaminated RGB (cleaned from field and asymptotic giant-branch stars) and determine its width to ascertain and quantify any intrinsic broadening. We hence defined our sample of RGB stars using an equivalent approach to that used by \cite[][]{Zhang2018}, defining RGB selection parallellograms in a combination of three different CMDs, including $(m_{\rm F475W} - m_{\rm F814W})$ versus $m_{\rm F814W}$, $(m_{\rm F336W} - m_{\rm F475W})$ versus $m_{\rm F475W}$, and $(m_{\rm F336W} - m_{\rm F814W})$ versus $m_{\rm F814W}$. Our final sample covers the range $15.0 \leq m_{\rm F814W} \leq 19.9$ mag. We next imposed magnitude and color cuts near the red clump, including only those stars with luminosities equal to or brighter and redder than the red clump. This removed some bluer stars from our sample. Figure \ref{fig1} (left) displays the $(m_{\rm F475W} - m_{\rm F814W})$ versus $m_{\rm F814W}$ CMD of NGC 2173 with the final selection of RGB stars marked (red solid circles).

Figure \ref{fig1} shows that the cluster's TRGB appears to be bifurcated. If this is an intrinsic property, it would lend strong support to the notion that NGC 2173 is the youngest cluster known with evidence of MPs. This feature was not affected by our decontamination procedure. We will next discuss how we quantified the cluster's RGB bimodality, its pseudo-CMD, and a pseudo-$UBI$ diagram generated using UV, visible, and infrared {\sl HST} photometry.

\section{ANALYSIS AND RESULTS}
\label{results.sec}

We compared our RGB stars with the Padova stellar evolution models \citep[PARSEC v.3.1;][]{Bressan2012PARSEC:Code}), and also with the MIST \citep{Choi2016} and BaSTI \citep[a Bag of Stellar Tracks and Isochrones;][]{PietrinferniADISTRIBUTION} isochrones, to obtain precise estimates of the cluster's age ($t$), metallicity ($Z$), distance modulus, and extinction. We determined $\log(t \mbox{ yr}^{-1}) = 9.13 \pm 0.02$ and $Z= 0.008 \pm 0.002$. 

We next explored whether MPs might be present among the cluster's RGB stars, particularly at the split TRGB. Figure \ref{fig1} (left, inset) shows the bifurcation in detail. We defined a central fiducial line in the $(m_{\rm F475W} - m_{\rm F814W}$) versus $m_{\rm F814W}$ CMD (blue solid line). We then calculated the separation, $\Delta_{\rm F485W,F814W}$, in $(m_{\rm F475W} - m_{\rm F814W}$) color for each RGB star from this fiducial line; see Figure \ref{fig1} (right). 
\begin{figure}
    \begin{center}
    \includegraphics[width=0.45\textwidth]{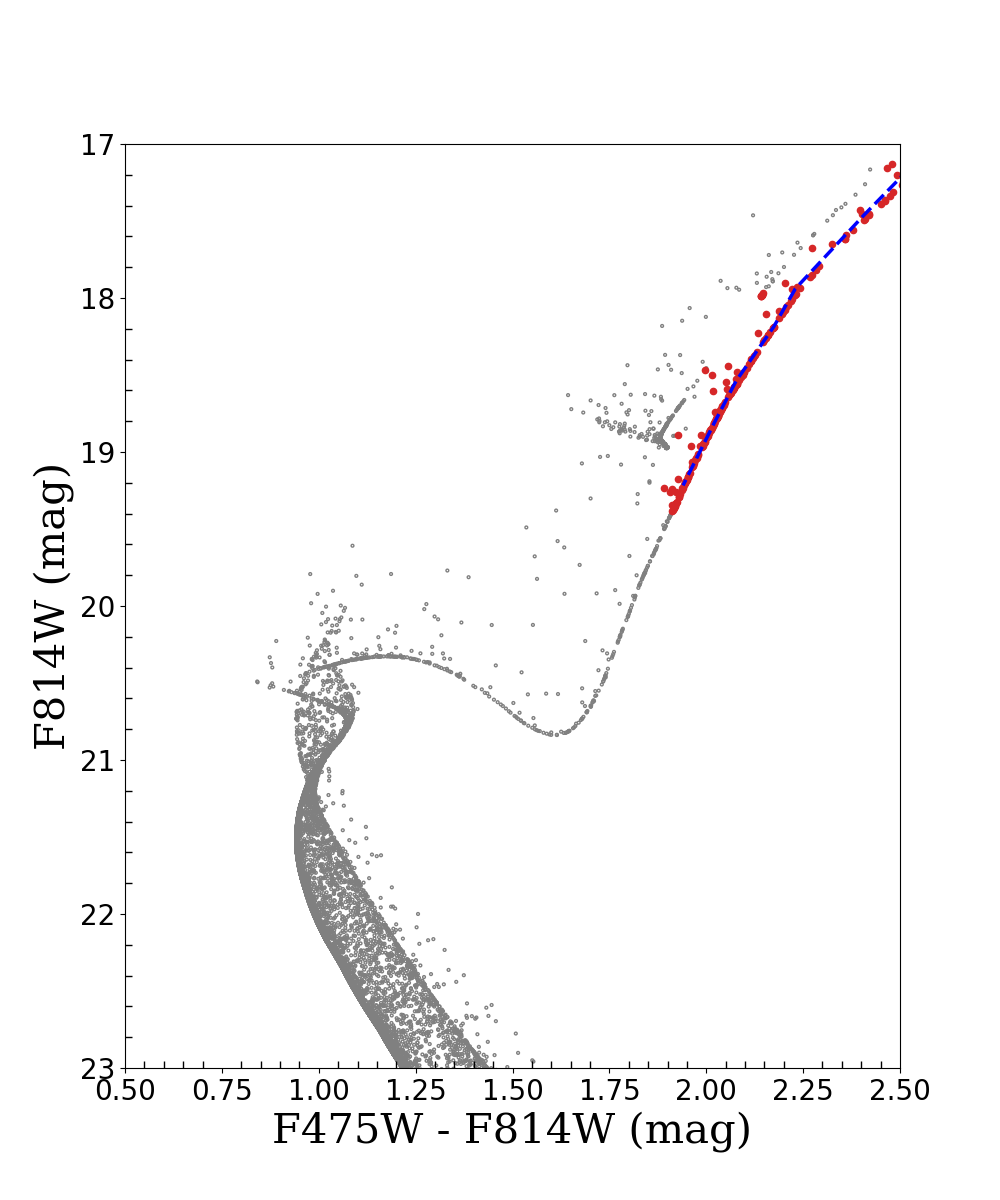}
    
    \includegraphics[width=0.45\textwidth]{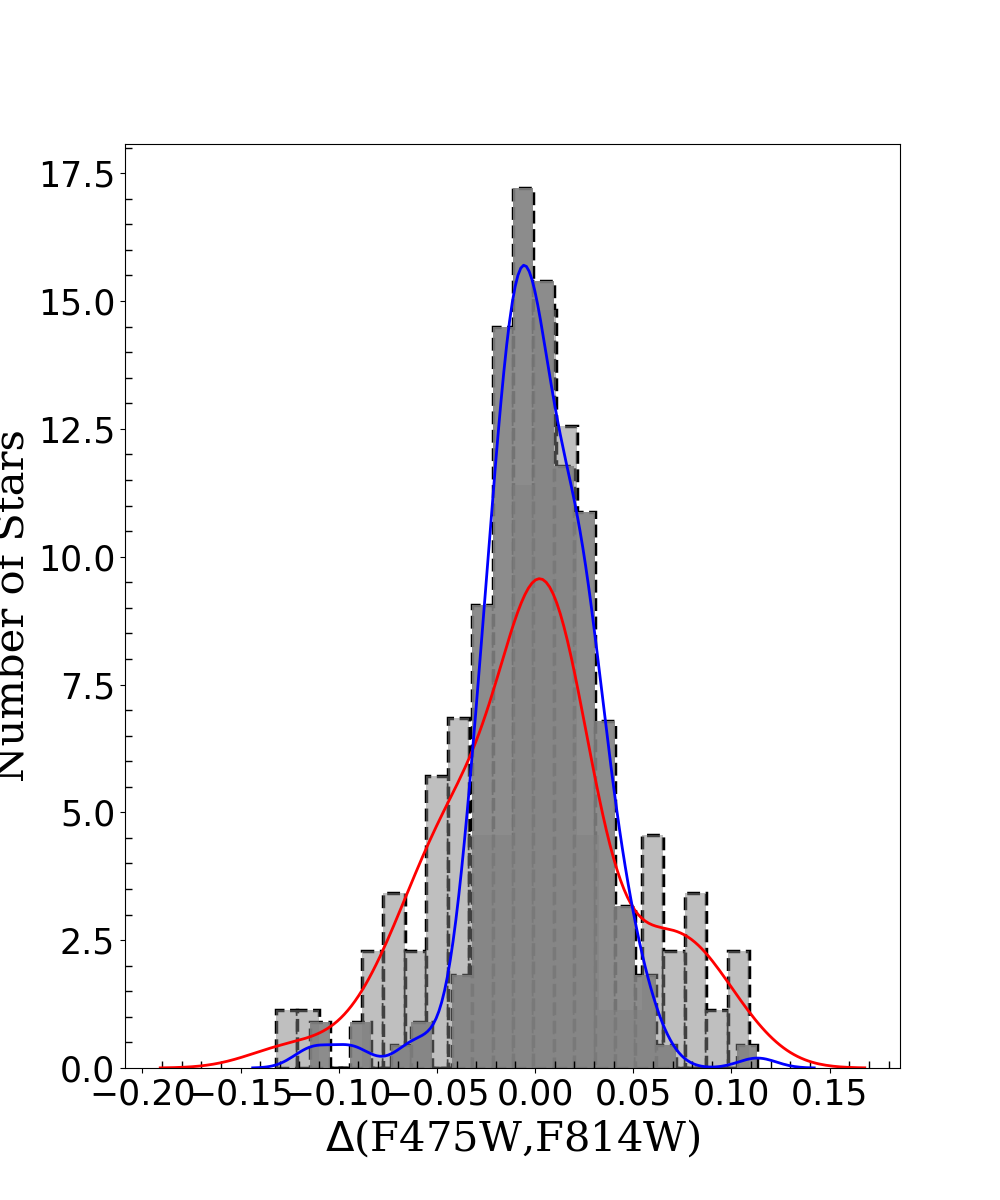}
    \end{center}
    \caption{(Top) Synthetic $(m_{\rm F336W} - m_{\rm F814W})$ versus $m_{\rm F814W}$ CMD of a cluster like NGC 2173, generated using SPISEA. The cluster's RGB, highlighted using red dots, does not show any indication of a bifurcated TRGB. (Bottom) Normalized histograms of observed and artificial RGB stars and their KDEs (light gray and red, and dark gray and blue, respectively).  }
    \label{fig2}
\end{figure}

We calculated the pseudo-color's mean and standard deviation ($\sigma$) and obtained the Gaussian probability density function (PDF); see Figure \ref{fig1} (right, yellow curve). To determine whether any additional peaks may have been smoothed out by the Gaussian PDF, we derived a kernel density estimation (KDE) using a Gaussian kernel; see Figure \ref{fig1} (right, solid red curve). The distribution's broadening and asymmetry can be seen clearly, along with a bump at $\Delta_{\rm F475W,F814W} = 0.06$ mag. We adopted the bump's color as our verticalized color to distinguish between the stellar populations; see Figure \ref{fig1} (right, black dashed line). 

\begin{figure}
    \begin{center}                                            
    \includegraphics[width=0.50\textwidth]{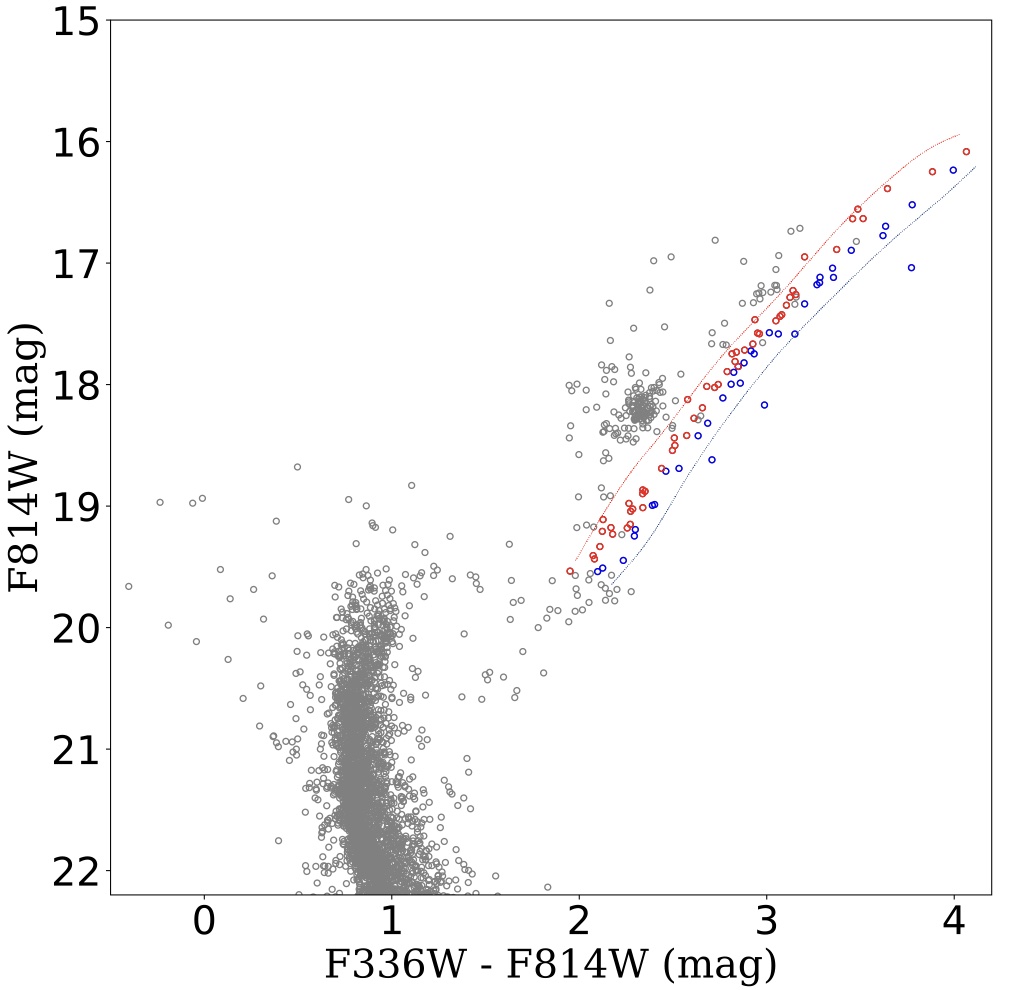}
    \end{center}
    \caption{($m_{\rm F336W} - m_{\rm F814W}$) versus $m_{\rm F814W}$ CMD of NGC 2173 obtained from the WFC3 catalogs. Blue, red solid lines: fiducial lines.}
    \label{fig3}
\end{figure}
\begin{figure*}
    \begin{center}
    \includegraphics[width=1\textwidth]{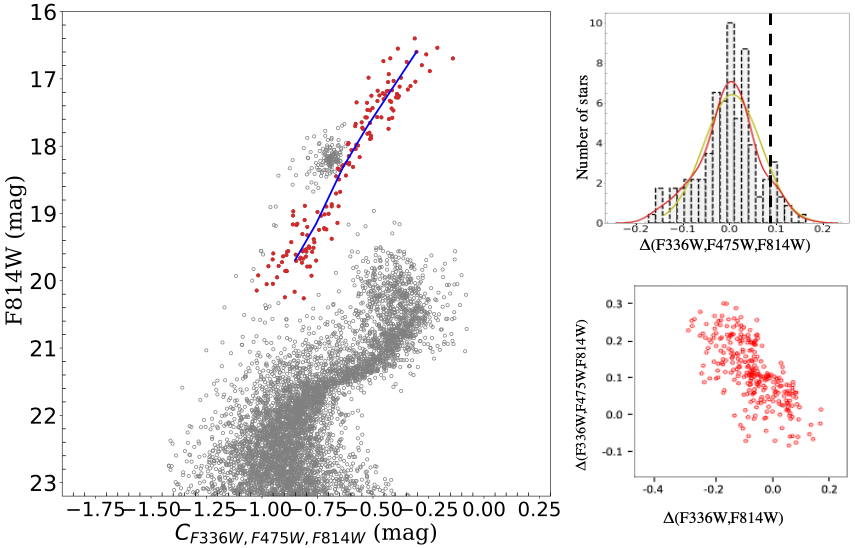}
   
    \end{center}
    \flushleft
    \caption{(Left) Selected RGB sample in the $m_{\rm F814W}$ versus $C_{\rm F336W,F475W,F814W}$ pseudo-CMD (red solid dots). Blue solid line: Fiducial line. (Top right) Verticalized RGB color distribution, $\Delta_{\rm F336W,F475W,F814W}$. Yellow, red solid lines: Gaussian PDF and corresponding KDE. Dashed line: Verticalized separation of the two distinct RGB populations. (Bottom right) Pseudo-$UBI$ diagram of NGC 2173 showing the verticalized color distribution ($\Delta_{\rm F336W,F814W}$) and the verticalized pseudo-color distribution ($\Delta_{\rm F336W,F475W,F814W}$). }
    \label{fig4}
\end{figure*}

We explored which external effects might have contributed to the cluster's broadened RGB. We first considered whether photometric uncertainties could have generated the bifurcation, performing artificial star tests using DOLPHOT to understand whether a significant number of stars could scatter randomly to resemble a bifurcated TRGB. To determine whether the RGB's measured width depends entirely on the photometric uncertainties or if any external parameters may affect its broadening and the observed bifurcation, we created artificial stellar catalogs containing 180,000 stars each. We used the {\tt -fakestar} option in DOLPHOT and reran DOLPHOT on the artificial stars, generating artificial CMDs for comparison purposes. We assigned random coordinates to the artificial stars based on our observed sample. To avoid overlaps between observed and artificial stars, we added random spatial offsets to the artificial stars. To finalize the RGB catalog, we randomly matched their coordinates with magnitudes obtained from the best-matching isochrone. The resulting artificial star catalog did not suggest any broadening or bifurcation of the RGB. We obtained the distribution of the observational uncertainties from our artificial star tests and compared its spread with that for our observed sample. Since the uncertainties from our artificial star tests are fairly large compared with the observational uncertainties, we rule out uncertainties as the dominant factor affecting the cluster's RGB broadening.  
 
The TRGB bifurcation in NGC 2173 is seen for RGB stars with $16.0 \le m_{\rm F814W} \le 17.2$ mag and $1.5 \le (m_{\rm F435W} - m_{\rm F814W}) \le 2.2$ mag (see Figure \ref{fig1}, right, inset). Scatter in the colors and their uncertainties could potentially create a bifurcation. However, the color uncertainties for the relevant $m_{\rm F814W}$ magnitude range are relatively small, $\sigma(m_{\rm F475W} - m_{\rm F814W}) \sim 0.002$ mag, whereas the typical color difference between both RGB branches is $\Delta (m_{\rm F475W} - m_{\rm F814W}) \approx 0.44$ mag for $16.0 \le m_{\rm F814W} \le 17.2$ mag, significantly larger than the typical photometric uncertainty. It is therefore unlikely that photometric uncertainties on their own could have yielded the observed split at the TRGB.

We also considered whether another cluster might be located along the line of sight. However, since the NGC 2173 CMDs do not exhibit multiple features anywhere else, we excluded that possibility. 

Finally, we explored extinction effects. We adopted a total extinction of $A_V = 0.15$ mag \citep{Li2018An}. Again using our artificial star sample, we tested extinction values ranging from $A_V = 0$ mag to $A_V = 0.15$ mag to explore the effects of fluctuations in the overall extinction, as well as the presence of differential extinction. We ruled out fluctuations in the overall extinction as the cause of the observed broadening and the bifurcation at the TRGB.

Differential reddening could potentially also cause a random broadening of the cluster's CMD features. To ascertain whether differential extinction could be responsible for the TRGB's bifurcation, we generated a synthetic cluster drawn from a simple stellar population using MIST isochrones and affected by variable extinction, using the Python package SPISEA \citep[Stellar Population Interface for Stellar Evolution and Atmospheres;][]{2020AJ....160..143H}.\footnote{https://spisea.readthedocs.io/en/latest/index.html} We repeated this process one thousand times, using variable extinction values across the cluster area to explore if a bifurcation developed in any of our runs, and how frequently. We did not detect any bifurcation at the TRGB, nor any broadening of the RGB, in any synthetic cluster (see Figure \ref{fig2}). To further investigate if the synthetic RGBs showed any indication of MPs, we selected our artificial RGB stars using the same process as for the observed RGB stars. Our final sample of artificial RGB stars is highlighted with red dots in Figure \ref{fig2} (top).

Figure \ref{fig2} (bottom) represents a comparison of the observed and artificial RGB stars. The light gray histogram displays the distribution of the observed RGB stars; the superimposed red solid line shows a clear bimodality. However, the dark gray histogram and the superimposed blue solid line do not show any bimodality of the artificial RGB stars. This demonstrates that the observed bimodality is real. In addition, the NGC 2173 CMD shows that the main RGB and SGB features are well-defined. Its red clump, at the level of the horizontal branch, is also compact. Our analysis thus implies (again) that differential extinction is indeed negligible. Therefore, we conclude that the observed bifurcation is most likely real. 

Next, we will characterize the cluster's two RGB populations. A combination of three filters commonly denoted as $C_{\rm index}$---specifically, $C_{\rm F336W, F475W, F814W} = (m_{\rm F336W} - m_{\rm F475W}) - (m_{\rm F475W} - m_{\rm F814W}$)---is a useful diagnostic to disentangle two populations characterized by different chemical abundance patterns \citep{Milone2012}. In addition, the MP phenomenon has recently been explored based on chromosome maps \citep{2015ApJ...808...51M, 2015MNRAS.447..927M}. The latter are pseudo-color--color plots that use specific combinations of UV (F275W), visible (F438W), and (infra-)red (F814W) filters. Briefly, we selected our RGB stars from the cluster's $(m_{\rm F475W} - m_{\rm F814W})$ versus $m_{\rm F814}$ CMD in the range $16.14 \le m_{\rm F814W} \le 19.94$ mag. Next, we created a pseudo-CMD of $m_{\rm F814W}$ versus $C_{\rm F336W,F475W,F814W}$, henceforth our `pseudo-$UBI$ diagram.' We used this latter diagram to define two fiducial lines (see Figure \ref{fig4}), equivalent to the 10th and 90th percentiles of the RGB distribution. 

\begin{figure}
    \begin{center}
    \includegraphics[width=0.48\textwidth]{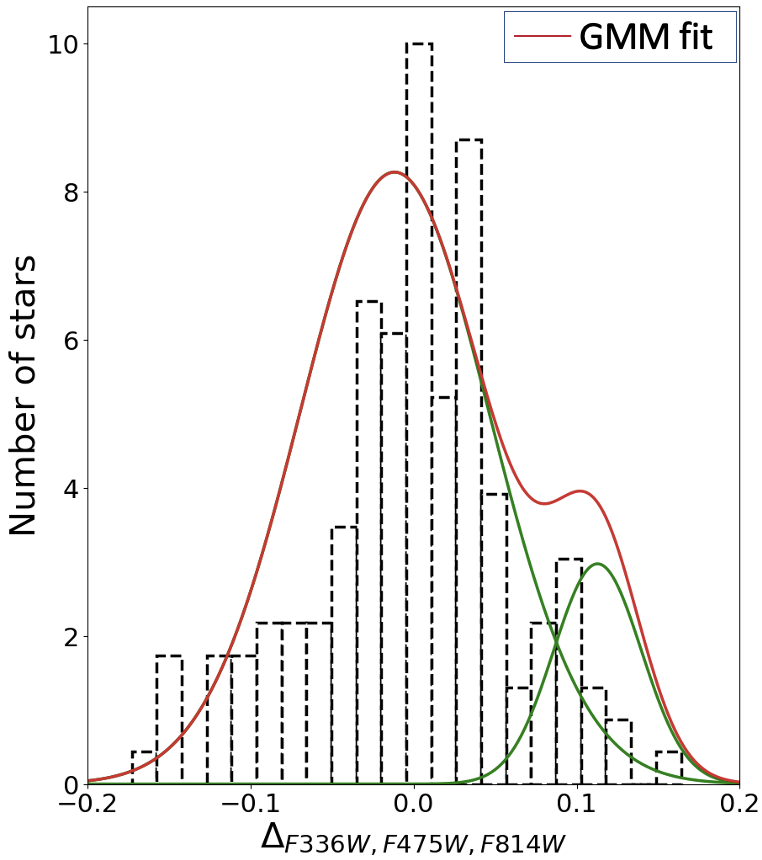}
    \end{center}
    \caption{Verticalized pseudo-color distribution, $\Delta_{\rm F336W,F475W,F814W}$, of the NGC 2173 RGB stars. Red solid line: Gaussian Mixture Model that fits the unbinned data best. Green: Individual Gaussian distributions.}
    \label{fig5}
\end{figure}

We consider these stars our true stellar sample for the generation of our pseudo-$UBI$ diagram. Next, we verticalized the RGB distribution following \citet{Milone2018}, creating $\Delta_{\rm F336W,F814W}$. We applied a similar approach to the pseudo-color diagram ($m_{\rm F814W}, C_{\rm F336W,F475W,F814W}$) to derive $\Delta_{\rm F336W,F475W,F814W}$. These values were used to compute and generate a ($\Delta_{\rm F336W,F814W}$, $\Delta_{\rm F336W,F475W,F814W}$) pseudo-$UBI$ diagram. We show the $\Delta_{\rm F336W,F475W,F814W}$ histogram in Figure \ref{fig4} (top right). We proceeded to fit a Gaussian PDF to these data and superimposed the corresponding KDE. The Gaussian PDF smoothed out a hint of bimodality in the $\Delta_{\rm F336W,F475W,F814W}$) distribution. Similar to the $\Delta_{\rm F475W,F814W}$ histogram (Figure \ref{fig1}), a bump is noticeable in the KDE at $\sim$0.07 mag. We used these values to generate a ($\Delta_{\rm F336W,F814W}, \Delta_{\rm F336W,F475W,F814W}$) pseudo-$UBI$ diagram for NGC 2173: see Figure \ref{fig4} (bottom right). 
\newpage
\section{DISCUSSION AND CONCLUSIONS}
We used high-resolution {\sl HST}/UVIS images in the F336W, F475W, and F814W filters to investigate the presence of MPs in NGC 2173, adopting one of the most powerful tools available at present, i.e., chromosome-like diagrams \citep{Milone2017}. These are pseudo-color diagrams that combine {\sl HST} UV, optical, and near-infrared photometric measurements to distinguish among stellar subpopulations with different chemical abundances. 

This is the first time that MPs have been identified in a cluster younger than NGC 1978 (age $\sim$2 Gyr) and also the first time a pseudo-$UBI$, chromosome-like diagram is presented for such a young cluster. Our UV--optical photometry, corrected for differential reddening, shows a distinctly broadened RGB. This broadening is clearly reflected in the bimodal distribution of the cluster's TRGB stars. This bimodality of NGC 2173 RGB stars has also been detected for the first time in a ($\Delta_{\rm F336W,F475W,F814W}, \Delta_{\rm F336W,F814W}$) pseudo-$UBI$ diagram, where the distribution of the RGB stars extends well beyond the photometric errors.

\citet{Martocchia2018AgeClusters} suggested that some mechanism operating only in stars with masses $\la 2 M_\odot$, the approximate mass of RGB stars at $t \sim 2$ Gyr, might be responsible for generating MPs. Typically, stars above this mass limit do not have strong magnetic fields, whereas stars below it do. This might be related to stellar rotation; only stars more massive than the mass limit show evidence of rapid rotation in the cluster CMD, particularly in their extended main-sequence turnoff (eMSTO) regions. NGC 2173 does not feature an eMSTO, so we can assume that rapid stellar rotation was not a major factor causing the bimodality. Other factors such as metallicity or age dispersions cannot play an important role either, given that we did not discern any stretch in magnitude anywhere else in the NGC 2173 CMDs. Therefore, light-element abundance variations among our cluster's RGB stars are the most likely explanation of the observations. The presence of light-element abundance variations is common for cluster MS and RGB stars less massive than $1.5 M_\odot$ \citep[][]{Martocchia2018a}; however, spectroscopic follow-up is required to confirm our hypothesis.

Spectroscopic analysis of five RGB stars in NGC 2173 was presented by 
\citet{Mucciarelli2008THE}. Those authors did not find any significant star-to-star elemental abundance variations. However, their sample of five stars was statistically too small. We showed that the RGB stars in NGC 2173 exhibit extended color and pseudo-color distributions with a hint of a bimodality. To verify whether this apparent bimodality can be best approximated by one or multiple Gaussian distributions, we applied Gaussian Mixture Modeling \citep[GMM;][]{Muratov2010MODELINGCLUSTERS} to our unbinned RGB sample using our $\Delta_{\rm F336W,F475W,814W}$ measurements. We fitted the RGB data with the SCIKIT-LEARN python package MIXTURE\footnote{http://scikit-learn.org/stable/modules/mixture.html} (Figure \ref{fig5}). This method applies an expectation-maximization algorithm for fitting multiple Gaussian models. Figure \ref{fig5} demonstrates that the NGC 2173 RGB data are best fitted by two Gaussian components. We categorize these two subpopulations as first- and second-generation stars based on the areas under the Gaussian profiles. GMM thus yielded $\frac{N_{\text{first generation}}}{N_{\text{total}}} = 0.64 \pm 0.03$ and $\frac{N_{\text{second generation}}}{N_{\text{total}}} = 0.36 \pm 0.03$, where $N_{\text{total}}$ is the total number of RGB stars.

An important implication of our discovery of MPs in the young, 1.7 Gyr-old cluster NGC 2173 is that it suggests that ancient Galactic GCs and young massive clusters might share a common formation process, since both types of clusters have now been confirmed to host MPs. Young clusters in the Magellanic Clouds are ideal laboratories for use as tracers of their host galaxies' evolution histories, given that they can be readily distinguished from the underlying field-star population at significant distances. Moreover, such massive clusters host statistically significant numbers of stars, which hence can be used as ideal proxies for stellar evolution analyses.

\section*{Acknowledgements}
SK acknowledges funding from the International Macquarie Research Excellence Scheme (iMQRES). DK acknowledges support from an Australian Research Council (ARC) Discovery Early Career Research Award (DECRA), grant DE190100813. This research was partially supported by the ARC Centre of Excellence for All Sky Astrophysics in 3 Dimensions (ASTRO3D) through project CE170100013. This paper is based on observations made with the NASA/ESA {\sl HST} (obtained from the Hubble Legacy Archive), a collaboration of the Space Telescope Science Institute (STScI/NASA), the Space Telescope European Coordinating Facility (ST-ECF/ESA), and the Canadian Astronomy Data Centre (CADC/NRC/CSA). We have made use of BaSTI web tools. 

\bibliography{NGC_2173.bib} 

\begin{thebibliography}{}
\expandafter\ifx\csname natexlab\endcsname\relax\def\natexlab#1{#1}\fi
\providecommand{\url}[1]{\href{#1}{#1}}

\bibitem[{Bastian {et~al.}(2013)Bastian, Cabrera-Ziri, Davies, \&
  Larsen}]{Bastian2013}
Bastian, N., Cabrera-Ziri, I., Davies, B., \& Larsen, S.~S. 2013, MNRAS, 436,
  2852

\bibitem[{Bressan {et~al.}(2012)Bressan, Marigo, Girardi, Salasnich, Dal~Cero,
  Rubele, \& Nanni}]{Bressan2012PARSEC:Code}
Bressan, A., Marigo, P., Girardi, L., {et~al.} 2012, MNRAS, 427, 127

\bibitem[{Carretta {et~al.}(2009)Carretta, Bragaglia, Gratton, D'Orazi,
  Lucatello, Carretta, Bragaglia, Gratton, D'Orazi, \&
  Lucatello}]{Carretta2009IntrinsicClusters}
Carretta, E., Bragaglia, A., Gratton, R., {et~al.} 2009, A{\&}A, 508, 695

\bibitem[{Choi {et~al.}(2016)Choi, Dotter, Conroy, Cantiello, Paxton, \&
  Johnson}]{Choi2016}
Choi, J., Dotter, A., Conroy, C., {et~al.} 2016, ApJ, 823, 102

\bibitem[{Dalessandro {et~al.}(2019)Dalessandro, Ferraro, Bastian, Cadelano,
  Lanzoni, \& Raso}]{Dalessandro2019TheContamination}
Dalessandro, E., Ferraro, F.~R., Bastian, N., {et~al.} 2019, A{\&}A, 621, A45

\bibitem[{{Dalessandro} {et~al.}(2016){Dalessandro}, {Lapenna}, {Mucciarelli},
  {Origlia}, {Ferraro}, \& {Lanzoni}}]{2016ApJ...829...77D}
{Dalessandro}, E., {Lapenna}, E., {Mucciarelli}, A., {et~al.} 2016, \apj, 829,
  77

\bibitem[{Decressin {et~al.}(2007)Decressin, Charbonnel, \&
  Meynet}]{Decressin2007OriginClusters}
Decressin, T., Charbonnel, C., \& Meynet, G. 2007, A{\&}A, 475, 859

\bibitem[{D'Ercole {et~al.}(2008)D'Ercole, Vesperini, D'Antona, McMillan, \&
  Recchi}]{DErcole2008FormationClusters}
D'Ercole, A., Vesperini, E., D'Antona, F., McMillan, S.~L., \& Recchi, S. 2008,
  MNRAS, 391, 825

\bibitem[{Dolphin(2011)}]{Dolphin2011DOLPHOT/WFC32.0}
Dolphin, A. 2011, {DOLPHOT/WFC3 User's Guide version 2.0}, , .
\newblock \url{http://americano.dolphinsim.com/dolphot/dolphotWFC3.pdf}

\bibitem[{Glatt {et~al.}(2008)Glatt, Grebel, Sabbi, Gallagher, Nota, Sirianni,
  Clementini, Tosi, Harbeck, Koch, Kayser, \& Da~Costa}]{Glatt2008}
Glatt, K., Grebel, E.~K., Sabbi, E., {et~al.} 2008, AJ, 136, 1703

\bibitem[{{Hosek} {et~al.}(2020){Hosek}, {Lu}, {Lam}, {Gautam}, {Lockhart},
  {Kim}, \& {Jia}}]{2020AJ....160..143H}
{Hosek}, Matthew~W., J., {Lu}, J.~R., {Lam}, C.~Y., {et~al.} 2020, \aj, 160,
  143

\bibitem[{Kapse {et~al.}(2021)Kapse, de Grijs, \&
  Zucker}]{Kapse2021Searching2213}
Kapse, S., de Grijs, R., \& Zucker, D.~B. 2021, MNRAS, 503, 6016

\bibitem[{Keller {et~al.}(2012)Keller, Mackey, Da~Costa, Keller, Mackey, \&
  Da~Costa}]{Keller2012Extended2209}
Keller, S.~C., Mackey, A.~D., Da~Costa, G.~S., {et~al.} 2012, ApJL, 761, L5

\bibitem[{Li {et~al.}(2018)Li, Deng, de~Grijs, Jiang, \& Xin}]{Li2018An}
Li, C., Deng, L., de~Grijs, R., Jiang, D., \& Xin, Y. 2018, ApJ, 856, 25

\bibitem[{{Li} {et~al.}(2019){Li}, {Sun}, {Hong}, {Deng}, {de Grijs}, \&
  {Sills}}]{2019ApJ...871..171L}
{Li}, C., {Sun}, W., {Hong}, J., {et~al.} 2019, \apj, 871, 171

\bibitem[{Marino {et~al.}(2008)Marino, Villanova, Piotto, Milone, Momany,
  Bedin, \& Medling}]{Marino2008}
Marino, A.~F., Villanova, S., Piotto, G., {et~al.} 2008, A{\&}A, 490, 625

\bibitem[{Marino {et~al.}(2016)Marino, Milone, Casagrande, Collet, Dotter,
  Johnson, Lind, Bedin, Jerjen, Aparicio, \&
  Sbordone}]{Marino2016ChemicalComponent}
Marino, A.~F., Milone, A.~P., Casagrande, L., {et~al.} 2016, MNRAS, 459, 610

\bibitem[{Martocchia {et~al.}(2017)Martocchia, Bastian, Usher,
  Kozhurina-Platais, Niederhofer, Cabrera-Ziri, Dalessandro, Hollyhead,
  Kacharov, Lardo, Larsen, Mucciarelli, Platais, Salaris, Cordero, Geisler,
  Hilker, Li, \& Mackey}]{Martocchia2017a}
Martocchia, S., Bastian, N., Usher, C., {et~al.} 2017, MNRAS, 468, 3150

\bibitem[{Martocchia {et~al.}(2018{\natexlab{a}})Martocchia, Cabrera-Ziri,
  Lardo, Dalessandro, Bastian, Kozhurina-Platais, Usher, Niederhofer, Cordero,
  Geisler, Hollyhead, Kacharov, Larsen, Li, Mackey, Hilker, Mucciarelli,
  Platais, \& Salaris}]{Martocchia2018AgeClusters}
Martocchia, S., Cabrera-Ziri, I., Lardo, C., {et~al.} 2018{\natexlab{a}},
  MNRAS, 473, 2688

\bibitem[{Martocchia {et~al.}(2018{\natexlab{b}})Martocchia, Niederhofer,
  Dalessandro, Bastian, Kacharov, Usher, Cabrera-Ziri, Lardo, Cassisi, Geisler,
  Hilker, Hollyhead, Kozhurina-Platais, Larsen, Mackey, Mucciarelli, Platais,
  \& Salaris}]{Martocchia2018a}
Martocchia, S., Niederhofer, F., Dalessandro, E., {et~al.} 2018{\natexlab{b}},
  MNRAS, 4705, 4696

\bibitem[{{Milone} {et~al.}(2009){Milone}, {Bedin}, {Piotto}, \&
  {Anderson}}]{2009A&A...497..755M}
{Milone}, A.~P., {Bedin}, L.~R., {Piotto}, G., \& {Anderson}, J. 2009, A{\&}A,
  497, 755

\bibitem[{Milone {et~al.}(2016)Milone, Marino, D'Antona, Bedin, Da~Costa,
  Jerjen, \& Mackey}]{Milone2016a}
Milone, A.~P., Marino, A.~F., D'Antona, F., {et~al.} 2016, MNRAS, 458, 4368

\bibitem[{Milone {et~al.}(2012)Milone, Piotto, Bedin, Aparicio, Anderson,
  Sarajedini, Marino, Moretti, Davies, Chaboyer, Dotter, Hempel,
  Mar{\'{i}}n-Franch, Majewski, Paust, Reid, Rosenberg, \& Siegel}]{Milone2012}
Milone, A.~P., Piotto, G., Bedin, L.~R., {et~al.} 2012, A{\&}A, 540, A16

\bibitem[{{Milone} {et~al.}(2015{\natexlab{a}}){Milone}, {Marino}, {Piotto},
  {Renzini}, {Bedin}, {Anderson}, {Cassisi}, {D'Antona}, {Bellini}, {Jerjen},
  {Pietrinferni}, \& {Ventura}}]{2015ApJ...808...51M}
{Milone}, A.~P., {Marino}, A.~F., {Piotto}, G., {et~al.} 2015{\natexlab{a}},
  \apj, 808, 51

\bibitem[{{Milone} {et~al.}(2015{\natexlab{b}}){Milone}, {Marino}, {Piotto},
  {Bedin}, {Anderson}, {Renzini}, {King}, {Bellini}, {Brown}, {Cassisi},
  {D'Antona}, {Jerjen}, {Nardiello}, {Salaris}, {Marel}, {Vesperini}, {Yong},
  {Aparicio}, {Sarajedini}, \& {Zoccali}}]{2015MNRAS.447..927M}
---. 2015{\natexlab{b}}, \mnras, 447, 927

\bibitem[{Milone {et~al.}(2017)Milone, Marino, D'Antona, Bedin, Piotto, Jerjen,
  Anderson, Dotter, Di~Criscienzo, \& Lagioia}]{Milone2017}
Milone, A.~P., Marino, A.~F., D'Antona, F., {et~al.} 2017, MNRAS, 465, 4363

\bibitem[{Milone {et~al.}(2018)Milone, Marino, Renzini, D'Antona, Anderson,
  Barbuy, Bedin, Bellini, Brown, Cassisi, Cordoni, Lagioia, Nardiello,
  Ortolani, Piotto, Sarajedini, Tailo, van~der Marel, \&
  Vesperini}]{Milone2018}
Milone, A.~P., Marino, A.~F., Renzini, A., {et~al.} 2018, MNRAS, 481, 5098

\bibitem[{Mucciarelli {et~al.}(2008)Mucciarelli, Carretta, Origlia, \&
  Ferraro}]{Mucciarelli2008THE}
Mucciarelli, A., Carretta, E., Origlia, L., \& Ferraro, F.~R. 2008, AJ, 136,
  375

\bibitem[{Mucciarelli {et~al.}(2010)Mucciarelli, Origlia, \&
  Ferraro}]{Mucciarelli2010ChemicalCloud}
Mucciarelli, A., Origlia, L., \& Ferraro, F.~R. 2010, ApJ, 717, 277

\bibitem[{Muratov \& Gnedin(2010)}]{Muratov2010MODELINGCLUSTERS}
Muratov, A.~L., \& Gnedin, O.~Y. 2010, ApJ, 718, 1266

\bibitem[{Niederhofer {et~al.}(2017)Niederhofer, Bastian, Kozhurina-Platais,
  Larsen, Hollyhead, Lardo, Cabrera-Ziri, Kacharov, Platais, Salaris, Cordero,
  Dalessandro, Geisler, Hilker, Li, Mackey, \& Mucciarelli}]{Niederhofer2017b}
Niederhofer, F., Bastian, N., Kozhurina-Platais, V., {et~al.} 2017, MNRAS, 465,
  4159

\bibitem[{Pietrinferni {et~al.}(2006)Pietrinferni, Cassisi, Salaris, \&
  Castelli}]{PietrinferniADISTRIBUTION}
Pietrinferni, A., Cassisi, S., Salaris, M., \& Castelli. 2006, ApJ

\bibitem[{Piotto {et~al.}(2015)Piotto, Milone, Bedin, Anderson, King, Marino,
  Nardiello, Aparicio, Barbuy, Bellini, Brown, Cassisi, Cool, Cunial,
  Dalessandro, D'Antona, Ferraro, Hidalgo, Lanzoni, Monelli, Ortolani, Renzini,
  Salaris, Sarajedini, Van Der~Marel, Vesperini, \&
  Zoccali}]{Piotto2015ThePopulations}
Piotto, G., Milone, A.~P., Bedin, L.~R., {et~al.} 2015, AJ, 149, 91

\bibitem[{Zhang {et~al.}(2018)Zhang, de~Grijs, Li, \& Wu}]{Zhang2018}
Zhang, H., de~Grijs, R., Li, C., \& Wu, X. 2018, ApJ, 853, 186

\end{thebibliography}



\end{document}